# ARTICLE

# Single particle tracking of polymer aggregates inside disordered porous media

Yusaku Abe,*[a] Naoki Tomioka [a] and Yu Matsuda *[a]

The diffusion motions of individual polymer aggregates in disordered porous media were visualized using the single particle tracking (SPT) method because the motions inside porous media play important roles in various fields of science and engineering. The aggregates diffused on the surfaces of pores; continuous adsorption and desorption processes were observed. The relationship between the size of the aggregates and pore size was analysed based on diffusion coefficients, moment scaling spectrum (MSS) slope, and diffusion anisotropy analyses. The obtained diffusion coefficients were different for different the aggregate and pore sizes. The MSS Slope analysis indicated that more than 85% of the aggregates were confined diffusion for all the conditions investigated. The diffusion anisotropies analysis suggested that the diffusions of the aggregates were anisotropic. The interactions between the aggregates and the pores were complex and exhibited different motions than surface diffusion of smooth surfaces.

## Introduction

Porous media have been widely used in industry because of their unique properties, which are caused by geometrical and surface interaction.[1] For example, the porous media have been used as catalysts,[2] adsorbents,[3,4] and filters.[5] In polymer science, porous media are also powerful materials to catalysts,[6] filters,[7] and adsorbents of polymers.[8] Since polymers tend to aggregate in solution,[9–11] the motion of polymers inside porous media depends not only on the properties of single polymer chain but also on those of polymer aggregate.[12] Therefore, the understand of the aggregate motions inside porous media is important for designing of desired polymers and porous media.

The polymer aggregates diffuse through interactions with the pores inside porous media. The interactions are assumed to vary with the size of the aggregates, the pore size, and the pore geometry. Sugar et al.,[13] revealed that the motion of the aggregates inside straight flow channels was affected by the size of the aggregates and the width of flow channels by fluorescence imaging of flowing fluorescent polymers and their aggregates. Even the interaction between such a simple geometry of the channel and the aggregates induced complex behaviours of the aggregates, which has led to an interest in the behaviours of polymer aggregates in disordered pores with more complex geometries.

We have investigated that the diffusion motion of nano-particles inside disordered porous media was anisotropic, and the one reason of this was the heterogeneous structure of the disordered porous media.[14] Miyagawa et al.,[15] investigated the polymer behaviour inside a single mesoporous silica particle by measuring the distribution of the fluorescence intensity from polymers and revealed that the motion of polymers depended on the pore size. Skaug et al.,[16] adopted single molecule tracking method to visualize the surface diffusion of polymers on a substrate. However, these studies have not measured and analysed the motion of the individual polymer aggregates inside disordered porous media. Visualization of the motion of the individual polymer aggregates are needed to reveal the features of the diffusion of the aggregates.

In the present study, we adopted single particle tracking (SPT) method[17–19] to investigate on polymer aggregate behaviour inside a monolithic silica column.[20] The SPT method is a powerful tool for investigating the motion of individual probes because it directly records the motion using a fluorescence microscope. Based on the trajectory data obtained by the SPT method, we can characterize the behaviour of the individual aggregates, such as diffusion coefficient, diffusion motion, and diffusion anisotropy. The monolithic silica columns used in this study had unimodal distribution of pore size. The diffusion motion of the polymer aggregates inside columns may be complex because the orientation, the pore size, and the shape inside columns are randomly arranged. In fact, our previous study[14] revealed that the nano-particle motion inside the monolithic column is heterogeneous. We investigated the effects of pore size and the aggregation size on the motion of the aggregate. The suppression of the diffusion coefficients and modes were investigated. The diffusion anisotropy was analysed by considering the displacement probability distributions (DPDs) and the relation between angle and displacement obtained from three consecutive steps of trajectory.

[a.] *Department of Modern Mechanical Engineering, Waseda University, 3-4-1 Ookubo, Shinhuku-ku, Tokyo, Japan.*
*Corresponding author: Y.A.: ya-jupiter0309@toki.waseda.jp and Y.M.: y.matsuda@waseda.jp

## Materials and Methods

**Sample Preparation**

We investigated diffusion motion of polymer aggregates in a silica monolithic column (Ex-Pure, Kyoto Monotech Co. Ltd., Japan) having disordered pore networks. Figs. 1a and 1b show Scanning electron microscope (SEM) images of the columns with pore size distribution peaks at 2 μm and 5 μm, respectively. As shown in Figs. 1a and 1b, the columns have disordered and continuous pores formed by polymerisation-induced phase separation.[20] The pore size distributions of the columns were measured by a mercury porosimeter (AutoPore IV 9500, Shmadzu Co. Ltd., Japan) as shown in Fig. 1c. The columns with small (2 μm) and large (5 μm) pores have 1.0-2.1 μm and 4.5-5.2 μm pores, respectively. The tortuosity factors of the columns were 1.58 and 1.51 for the columns with small (2 μm) and large (5 μm) pores, respectively. The porosity factors were 56.7% and 49.5% for the columns with small pores and large pores, respectively.

Luminescent poly(p-phenylenevinylene) (PPV) copolymer (Livilux®, PDY-132, Sigma-Aldrich, USA) was used as the probe polymer aggregate. We investigated the dependence of the size of the polymer aggregates on diffusion motion. Two different sizes of the aggregates were prepared from the $8.7 \times 10^{-3}$ wt% PPV copolymer stock solution in toluene (99.5% pure, Fujifilm Wako Pure Chemical Corporation, Japan): one was prepared by passing the stock solution through a 0.2 μm syringe filter (Membrane Solutions Co. Ltd., USA). The size of the obtained aggregates was measured as 20-150 nm, centred at 40 nm (see Fig. 1d), using a dynamic light scattering (DLS) analyser (nanoPartica SZ-100V2, HORIBA Co. Ltd., Japan). The other aggregate was prepared via a reprecipitation method. The stock solution (2.5 mL) was added to 1 mL of methanol (99.8% pure, Fujifilm Wako Pure Chemical Corporation, Japan). After filtration using a 5 μm syringe filter (Membrane Solutions Co. Ltd., USA), the aggregates of 80-450 nm, centred at 171 nm, were obtained by evaporation of the solvent (see Fig. 1d).

**SPT measurements**

We measured the aggregate motions in the column using the same SPT measurement system used in our previous studies.[14,21,22] Here, we provide a brief explanation of the measurement. The SPT measurements were conducted at room temperature of 25 °C. The monolithic silica column was placed in a custom-made glass cell (diameter: 2.9 mm, depth: 1.4 mm) made on a cover glass (Thickness No.1, Matsunami Glass Co. Ltd., Japan), and a dilute solution of the aggregates was poured into it. The fluorescence emitted from the aggregate was measured using an inverted fluorescence microscope (IX-73, Olympus Co. Ltd., Japan), a confocal scanner unit (CSU-X1, Yokogawa Electric Co. Ltd., Japan), and an oil-immersion objective lens of 100×, NA = 1.45, and WD = 0.13 mm (UPLXAPO100XO, Olympus Co. Ltd., Japan). The aggregates were photo-excited using a solid-state laser with an emission wavelength of 488 nm (OBIS488LS, Coherent CO. Ltd., USA). The output power of the laser was set at 135 mW. The fluorescence filtered by an optical bandpass filter (FF01-565/133-25, Semrock, USA) was captured using an electron-multiplying charge-coupled device (EMCCD) camera (C9100-23B, ImagEM X2, Hamamatsu Photonics Co. Ltd., Japan) at 50 frames per second (fps) (exposure time: 20 ms). In this condition, the one pixel of the obtained image corresponded to an actual length of 0.08 μm. We took images in several cross sections of 20 to 30 μm from the glass surface by changing focal plane using a piezo actuator (P-725K, Physik Instrumente GmbH & Co. KG, Germany). We conducted the SPT measurements under the following three combinations of conditions for the aggregates and pore sizes: (a) the aggregations with a diameter of $0.2d$ in the column with $8d$ pores, (b) the aggregations with a diameter of $0.2d$ in the column with $8d$ pores, and (c) t the aggregations with a diameter of $0.2d$ in the column with $8d$ pores, where $d$ is the diameter of the aggregates of 171 nm. Ten measurements were conducted for each condition, and 250 images were captured for each measurement.

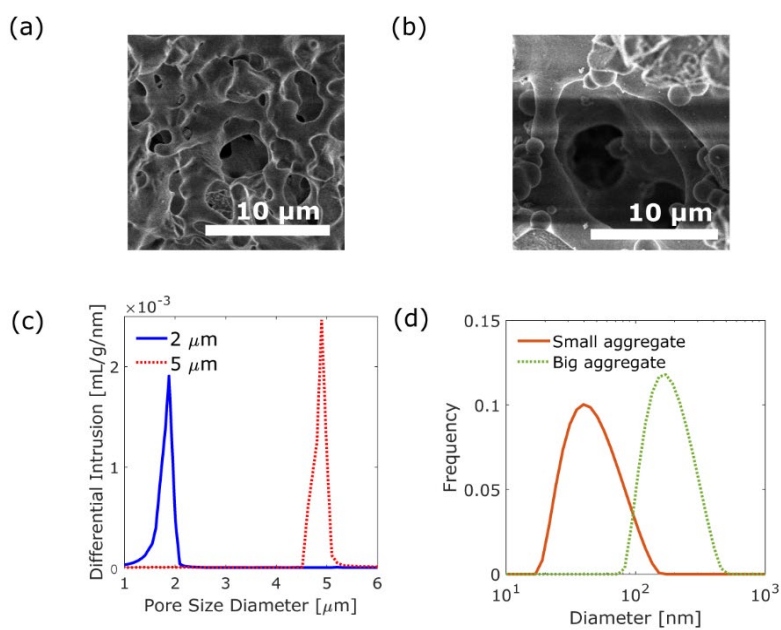

Fig. 1 SEM images of silica monolithic columns with pore size distribution peak at 2 μm (a) and 5 μm (b). (c) Pore size distributions of silica monolithic columns with pore size: the peak at 2 μm (solid blue line) and 5 μm (dashed red line). (d) Histogram of DLS of small aggregates with diameter distribution peak at 40 nm (solid orange line) and large aggregates with diameter distribution peak at 171 nm (dashed green line).

**Analysis of trajectories**

Time-series images of the aggregate motion were analysed using ImageJ Fiji[23] and PartcileTracker[24–26]. Each aggregate trajectory was extracted using ParticleTracker with the appropriate parameters. For images of small aggregates, the parameters were set as radius=5 px, cutoff=0.001, and per/abs=0.15. For images of large aggregates, parameters were set as radius=6 px, cutoff=0.001, and per/abs=0.05. The position vector of an aggregate at time step $t$ is represented by $\mathbf{X}(t)$, where $t$ (=0, 1, 2, …, $T$) represents time step and $T$ is the total time steps of the trajectory. For two-dimensional trajectory, the relation between the ensemble average of the $\nu$th moment of displacement and the lag time $\tau$ is expressed as

$$\langle \|\mathbf{X}(t+\tau) - \mathbf{X}(t)\|^\nu \rangle = 4D_\nu \tau^\gamma, \quad (1)$$

where $D_\nu$ is the generalized diffusion coefficient, $\gamma$ is a scaling coefficient, $\langle \bullet \rangle$ represents the ensembled average, and $\|\bullet\|$ represents the Euclidean norm. When $\nu = 2$ and $\gamma = 1$, eqn (1) becomes the relation between well-known mean squared displacement (MSD) and the regular diffusion coefficient $D_{\nu=2}$.

Hereafter, we denote $D = D_{\nu=2}$ to simplify the notation. The moment scaling spectrum (MSS),[24,27] the slope of which $S_{MSS}$ is an indicator of diffusion modes of each aggregate, was calculated from eqn (1). For the Brownian diffusion, $S_{MSS}$ is calculated as 0.5. In contrast, $S_{MSS}$ is calculated as $0 < S_{MSS} < 0.5$ for the confined diffusion. For super diffusion, $S_{MSS}$ is calculated as $0.5 < S_{MSS} < 1$. In the analysis, trajectories with fewer than 6 steps were not considered because short trajectories provided less reliable results. It is expected that adsorbed aggregates tend to $0 < S_{MSS} < 0.5$ and desorbed aggregates tend to $0.5 < S_{MSS} < 1$.

For a more detailed analysis, three consecutive steps of trajectory[28] was investigated to extract the diffusion anisotropy. When three consecutive positions of $\mathbf{X}(t-1)$, $\mathbf{X}(t)$, and $\mathbf{X}(t+1)$ are considered, the displacement between later two positions $\|\mathbf{X}(t+1) - \mathbf{X}(t)\|$ and the angle between the displacements $\|\mathbf{X}(t) - \mathbf{X}(t-1)\|$ and $\|\mathbf{X}(t+1) - \mathbf{X}(t)\|$ are plotted as a scatter diagram. The horizontal and vertical axes of the diagram are described as $\alpha$ and $\beta$, respectively. The shape of the scatter diagram evaluated by the radius of gyration depends on the diffusion anisotropy. Details of the explanation are provided in our previous studies.[14]

## Results and discussion

### Direct visualization of Individual polymer aggregates

The trajectories of the aggregates of 171 nm ($d$) coloured according to their diffusion coefficients are shown in Fig. 2, where a background image of the monolithic column with the pore size of 5 μm ($29d$) is a time averaged measurement image obtained by SPT. As shown in Fig. 2a, most trajectories were located on the surface of pores. The diffusion coefficients were different for each trajectory. Figs. 2b and 2c are magnified images. As shown in Fig. 2b, some aggregates were adsorbed in a certain area on the surface of the pore, while other aggregates diffused along the surface. Moreover, desorption from the surface was measured as illustrated in Figs. 2b and c.

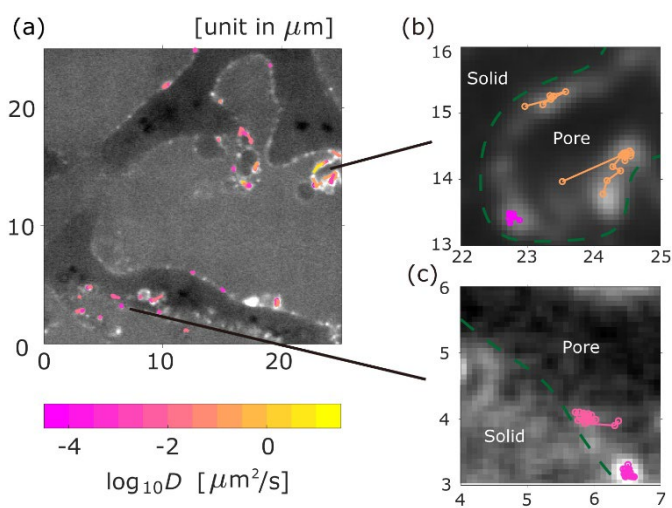

Fig. 2 Typical image of the measured trajectories, where the size of aggregates is 171.3 nm, and the pore size is 5 μm. Each trajectory is coloured by its diffusion coefficient $D$. (a) Trajectory distribution in area of 25 μm × 25 μm. (b), (c) Trajectories with adsorption and desorption. Pore walls are highlighted with dashed green line in (b) and (c).

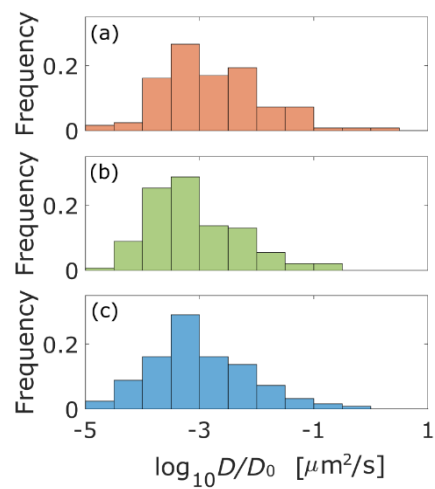

Fig. 3 Histograms of the diffusion coefficient, where the aggregates with diameter d for the columns with pore size 126d (a) and 8d (b) and the aggregates with diameter 0.2d for the column with pore size 8d (c). Diffusion coefficients of (a) and (b) are normalized by $D_0 = 4.7$ μm$^2$/s and (c) is normalized by $D_0 = 18.5$ μm$^2$/s.

**Diffusion coefficient and MSS slope analysis**

The statistical properties of the motion of the aggregates were investigated. Fig. 3 shows the histograms of diffusion coefficients. The experimental conditions of Figs. 3a and 3b are the aggregates of diameter 171 nm ($d$) inside the column of pore size $126d$ and $8d$, respectively. The experimental condition of Fig. 3c is the aggregates of diameter $0.2d$ inside the column of pore size $8d$. The numbers of trajectories analysed in Figs. 3a, b, and c are 1263, 1795, and 605, respectively. The diffusion coefficients were normalized by their bulk diffusion coefficients of 4.7 μm$^2$/s for the aggregates of the diameter $d$ and 18.5 μm$^2$/s for aggregates of the diameter $0.2d$, which were measured by the DLS analyser. Compared with the bulk diffusion coefficients, the average diffusion coefficients of these measurem000ents were 6.7% (Fig. 3a), 3.4% (Fig. 3b), and 1.1% (Fig. 3c). Comparing the results displayed in Fig. 3a and b, the diffusion motions in small pores were highly suppressed. In Figs. 3b and c, the larger the size of aggregates, the fewer aggregates with large diffusion coefficients.

The percentages of the aggregates which were $0 < S_{\text{MSS}} < 0.5$ are 87.3% (Fig. 4a), 87.6% (Fig. 4b), and 89.8% (Fig. 4c). These results indicate that diffusion motion of the most aggregates confined; namely, in conjunction with the visualization results in Fig. 2, most of the aggregates was the diffusion in the surface of the pore.

From Figs. 3 and 4, the diffusion coefficients decreased, and more aggregates were confined as the pore size decreased because the small pores on the surface of the columns would cause adsorption of the aggregates. Diffusion coefficients also decrease as the size of the aggregates increases due to the Stokes-Einstein law.[29]

**Diffusion anisotropy analysis**

The displacement of the aggregates for each consecutive time step defined as $\Delta \mathbf{X} = \|\mathbf{X}(t+1) - \mathbf{X}(t)\|$ also reflects the underlying motion. It is well known that the displacement probability distribution (DPD) follows a Gaussian distribution in Fickian diffusion, while tails of DPD is not described by Gaussian distribution but exponential in anomalous diffusion[30–33]. DPDs of the three experimental conditions are shown in Fig. 5, where $\Delta x$ and $G_S$ are respectively the displacement of aggregates along the horizontal axis at each step and frequency of this displacement and the black dotted line shows a Gaussian distribution. Comparing the two DPDs with different pore sizes, the probability of a large displacement with the large pore size is higher than that with the small pore size. On the other hand, compared with two DPDs of different size of aggregates, the distributions for $\Delta r < 2$ μm were similar. However, DPD of small aggregates is higher than DPD of large aggregates at $\Delta r > 2$ μm. These results indicate that the diffusion of the aggregates inside the disordered porous media is different from Brownian diffusion.

The aggregates diffusing along the surface of the pore show high anisotropy because these aggregates move in a specific direction; namely, along the surface structure. We introduced a $\alpha - \beta$ space and calculated from the position of real space to the position $\alpha - \beta$ space for aggregate trajectory, according to the proposed method in a previous study.[28] As shown in Figs. 6a-c, the aggregate positions of each time step in $\alpha - \beta$ space are plotted for each experimental condition. We also calculated radii of gyrations[28,34] for the scatter plot to evaluate the diffusion anisotropy. The coloured solid lines in Figs. 6a-c are the radius of gyrations of each experimental condition. The black solid lines in Figs. 6a-c are the circles which radii are equal to the radius of gyrations of the α axis. The ratios of the radius of the α axis versus the radius of the β axis/direction are 0.68 (Fig. 6a), 0.74 (Fig. 6b), and 0.83 (Fig. 6c). The result revealed polymer aggregates diffuse along the surface of the pore because these ratios are under 1 at all experimental conditions. Fig. 6d is the overlayed plot of the radii of gyrations in Fig. 6a-c. Compared with three radii, it is confirmed that displacements decrease as the size of the aggregate decreases, and pore size increases. As shown in Figs. 6a and 6b, the diffusion anisotropy decreases as the pore size decreases. This result indicates that the aggregates tend to diffuse inside small cavity or ledge on the surface when they were large relative to the size of the aggregates. The diffusion anisotropy also decreases as the size of the aggregates decreases. There are possible reasons as follows: firstly, the decrease in intermolecular forces due to decrease in the size of the aggregates triggers the desorption of the aggregates.[35] Secondly, the roughness of the surface increases relatively as the size of the aggregates decreases.[36] Finally, adhesion force between the aggregates and the surface decreases due to changes of polymer conformation, and the conformation triggers the desorption of the aggregates.[16]

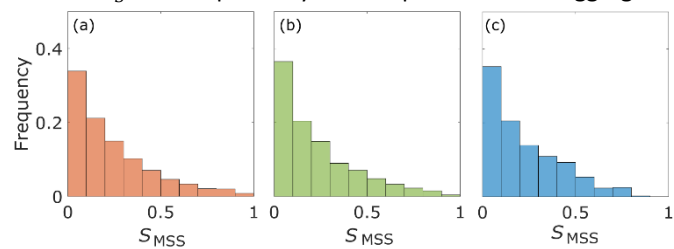

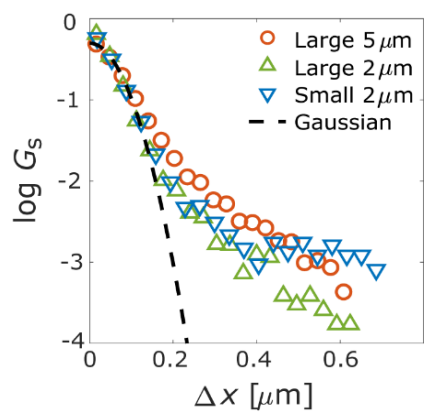

Fig. 4 Histograms of the MSS Slope $S_{\text{MSS}}$, where the aggregates with diameter d for the columns with pore size 126d (a) and 8d (b) and the aggregates with diameter 0.2d for the column with pore size 8d (c).

Fig. 5 Displacement probability distribution (DPD) of the aggregates with diameter d for the columns with pore size 126d (orange circle) and 8d (green triangle) and the aggregates with diameter 0.2d for the column with pore size 8d (blue inverted triangle).

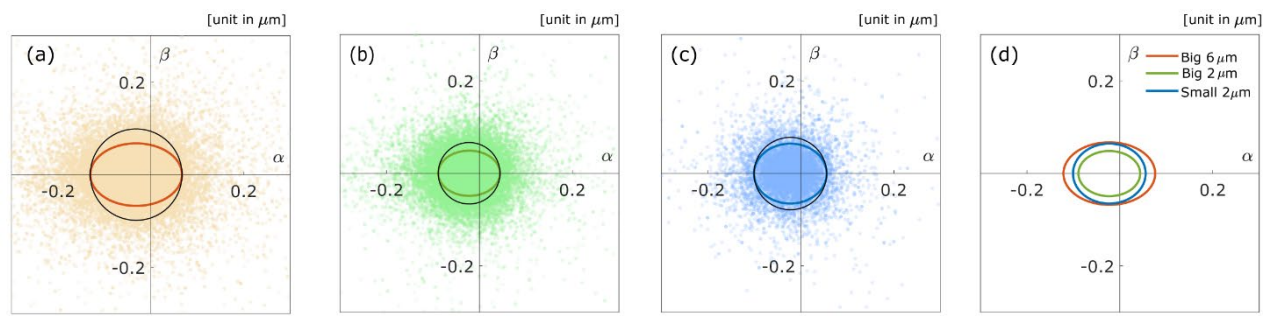

Fig. 6 Scatter diagrams for diffusion anisotropy for the aggregates with diameter d for the columns with pore size 126d (orange) and 8d (green) and the aggregates with diameter 0.2d for the column with pore size 8d (blue). The coloured line of ellipses are radii of gyrations, and the black solid lines are the circles which radii are equal to the radius of gyrations of the α axis. (d) The coloured lines of ellipses in (a)-(c) are overlayed.

## Conclusions

We investigated diffusion of polymer aggregates inside silica monolithic column by the SPT method. By visualization of each trajectory obtained from the SPT method, it was revealed that trajectories of each polymer aggregate located on the surface of pores. Statistical analysis of trajectories clarified that the distributions of diffusion coefficients are different for each experimental condition. Diffusion coefficients decrease as the size of the aggregates decreases, and the pore size increases. DPD also suggests that the difference of aggregate behaviour. Diffusion anisotropy described the probability of aggregates moving along the surface of the pore. We found that changes of adsorption property of aggregates depending on the changes of the size of the aggregates and pore size. This study visualizes polymer aggregates diffuse heterogeneously inside disordered porous media. Results of this study suggest that interactions between the aggregates and the pores differ depending on the size of the aggregates, the pore size, and the position of the aggregates on the surface.

## Data availability

The data supporting this article have been included in the manuscript or Supplementary Information. The code of "ParticleTracker" used to extracting aggregates from images can be found at:
https://sbalzarini-lab.org/?q=downloads/imageJ.
The version of the code employed for this study is version Version Nov 2016.

## Author contributions

Yusaku Abe: conceptualization, data curation, funding acquisition, formal analysis, investigation, methodology, software, validation, visualization, writing – original draft, and writing – review & editing. Naoki Tomioka: data curation, investigation, methodology, and software. Yu Matsuda: conceptualization, data curation, funding acquisition, investigation, methodology, project administration, supervision, and writing – review & editing.

## Conflicts of interest

There are no conflicts to declare.

## Acknowledgements

The authors thank Y. Okamura for assistance with SPT measurements. This work was partially supported by the Sasakawa Scientific Research Grant from The Japan Science Society and JSPS KAKENHI grant numbers 22H01421 and 23K22692.